\documentclass[12pt]{article}
\usepackage{citesort}
\usepackage{rotate}
\input{epsf}

\newcommand{\sect}[1]{\setcounter{equation}{0}\section{#1}}

\begin{document}

\title{Open inflation from non-singular instantons: \\
       Wrapping the universe with a membrane}
\author{{\sc Raphael Bousso}$^{\rm a}$\thanks{\it
 bousso@stanford.edu} \ 
  and {\sc Andrew Chamblin}$^{\rm b}$\thanks{\it
 hac1002@damtp.cam.ac.uk}
      \\[1 ex] {\it {\small
\begin{tabular}{ll}
$\!\!\!^{\rm a}$Department of Physics
& $\!\!\!^{\rm b}$DAMTP \\
Stanford University
& University of Cambridge   \\
Stanford, CA 94305-4060 \makebox[3em]{}
& Silver Street, Cambridge CB3 9EW \\
U.S.A.
& United Kingdom \\
\end{tabular} } }
       }
\date{DAMTP/R-98/23~~~~SU-ITP-98-25~~~~hep-th/9805167}

\maketitle

\begin{abstract}
  
  The four-form field recently considered by Hawking and Turok couples
  naturally to a charged membrane, across which the effective
  cosmological constant has a discontinuity. We present instantons for
  the creation of an open inflationary universe surrounded by a
  membrane.  They can also be used to describe the nucleation of a
  membrane on a pre-existing inflationary background. This process
  typically decreases the value of the effective cosmological constant
  and may lead to a novel scenario of eternal inflation.  Moreover, by
  coupling the inflaton field to the membrane, the troublesome
  singularities which arise in the Hawking-Turok model can be
  eliminated.

\end{abstract}


\pagebreak

\sect{Introduction: The Hawking-Turok model}

\subsection{Wave function and gravitational instantons}

As the recent developments in quantum cosmology have revived some
long-standing disputes about the foundations of the field, it is
appropriate for us to begin by stating our position on these questions
as far as they relate to the work in this paper.

A gravitational instanton is a Euclidean solution of the Einstein
equations with appropriate matter sources, and, in the case of
cosmology, an (effective) cosmological constant. It can typically be
applied in two ways. Within the wavefunction approach to quantum
cosmology~\cite{HarHaw83,Lin84b,Vil86}, it describes semiclassically
the creation of the universe from nothing. The corresponding
probability measure is obtained as the square of the amplitude of the
wavefunction.

The amplitude of the Hartle-Hawking~\cite{HarHaw83} wavefunction is
given semiclassically by the exponential of {\em minus} the Euclidean
instanton action. This wavefunction has the advantage that it is
well-defined beyond minisuperspace, and mathematically simple; it
merely paraphrases the rule that the likelihood of any state is
proportional to the exponential of its total entropy.  Precisely for
this reason, however, it predicts that the inflaton field should start
at the minimum of its potential, excluding any possibility of
sufficient inflation. Even after the anthropic principle is applied,
the predictions for $\Omega_0$ are unacceptable~\cite{HawTur98,Lin98}.

This problem does not occur if Linde's~\cite{Lin84b} or
Vilenkin's~\cite{Vil86} wave functions are used. While they are not
equivalent~\cite{Vil98b}, they both effectively yield a probability
measure of
\begin{equation}
P_{\rm L} = P_{\rm V} = 1/P_{\rm HH},
\end{equation}
giving greatest weight to high values of the inflaton field. This
typically leads to a long period of inflation yielding $\Omega_0 = 1$,
which is consistent with recent estimates of $\Omega_{\rm matter} +
\Omega_{\Lambda}$.

Linde's wavefunction was proposed as a tool to describe the instant of
the creation of the universe, when the only relevant degree of freedom
to be quantized was the scale factor. It cannot be generalized
straightforwardly to include other degrees of freedom, since they
would require the opposite Wick-rotation for
stability~\cite{Lin98,HawTur98b}. While Vilenkin's wave function is
perturbatively stable, it predicts that the universe is more likely to
start with a pair of black holes than
without~\cite{BouHaw95,BouHaw96}. Claims that Vilenkin's proposal is
``inapplicable'' to this question~\cite{GarVil96} would seem to imply
that his wave function is not sufficiently general to be of any
fundamental interest.

Thus all three proposals for the wave function of the universe suffer
from drawbacks; no single one of them answers all relevant questions
satisfactorily. Yet they have proven powerful tools for investigating
the initial conditions of the universe and other aspects of quantum
gravity. Keeping in mind that none of them are rigorously derived from
first principles, and that they may therefore provide answers to very
different questions, it seems entirely appropriate to proceed with
caution, and to apply each proposal only where it yields physically
reasonable results.  Independently of any probabilities assigned,
there is little dispute that cosmological instantons correspond to the
initial states of the universe that are semiclassically allowed.

The second application of gravitational instantons is far less
controversial: the description of non-perturbative fluctuations on a
pre-existing
background~\cite{ColDel80,GinPer83,Gib86,GarVil92,BouHaw95}.  To
obtain the semiclassical creation rate, $\Gamma$, for objects like
black holes, domain walls, cosmic strings, or simply true vacua, one
finds instantons that continue analytically to a Lorentzian universe
containing the desired features. Its action is calculated, and then
normalized by subtracting the action of the background instanton. (For
cosmological instantons this subtraction is implicit in the
construction of the saddlepoint path; see Ref.~\cite{BouCha97}.)  Then
the creation rate is given by
\begin{equation}
\Gamma = \exp \left[ -(I_{\rm instanton} - I_{\rm background})
\right].
\label{eq-pcr}
\end{equation}
Note that in this formula, as in all equations below, we take $I$ to
be the action of the full Euclidean solution, rather than the
half-bounce that is actually used in the interpolating path; this will
avoid many confusing factors of $2$.

\subsection{Action}

The Hawking-Turok~\cite{HawTur98c} (HT) model is Einstein gravity with
a scalar field $\phi$ that has a generic effective potential of
chaotic inflation, with a minimum at $\phi=0$, and no other stationary
points.  They also include a four-form field,
\begin{equation}
F_{\mu\nu\rho\lambda} = \partial_{[\mu} A_{\nu\rho\lambda]},
\label{eq-fa}
\end{equation}
which (unlike the inflaton) occurs naturally in 11D
supergravity~\cite{DufNil86}. It has no classical dynamics in four
dimensions, but can be used to cancel an effective cosmological
constant arising from the combined effects of other fields in the
theory. The Euclidean action is
\begin{eqnarray}
I_{\rm ht} & = & \int_M d^4\!x\, \sqrt{g}
    \left[ - \frac{R}{16 \pi G} + \frac{1}{2}(\partial \phi)^2
    + V(\phi) - \frac{1}{48} F_{\mu\nu\rho\lambda}
    F^{\mu\nu\rho\lambda} \right] \nonumber \\
    & & - \frac{1}{8 \pi G} \int_{\partial M} d^3\!x\, \sqrt{h} K
        + \frac{1}{24} \int_{\partial M} d^3\!x\, \sqrt{h}
          F^{\mu\nu\rho\lambda} n_\mu A_{\nu\rho\lambda},
\label{eq-action-ht}
\end{eqnarray}
where $K$ is the trace of the second fundamental form, $K_{ij}$ on any
boundary with unit normal vector $n_\mu$. The sign for the $F^2$ term
differs from that used in Refs.~\cite{BroTei87,BroTei88,DunJen90}, who
use, in the 4D effective action, the same sign that occurs in the
original 11D supergravity Lagrangian, so that the four-form field
gives a positive cosmological constant. In the 11D theory, however,
the four-form field makes a negative contribution towards the
effective cosmological constant in the 4D subspace. In order to
reproduce this from a 4D effective action, the sign of the $F^2$ term
must be changed~\cite{FreRub80}.  The first boundary term is the usual
Gibbons-Hawking term~\cite{GibHaw77b}. The second boundary term must
be included to obtain stationary action under variations that leave
$F$ fixed on the boundary. On-shell its value is negative twice the
$F^2$ contribution in the volume term of the action.

\subsection{Four-form field and effective cosmological constant}

The four-form field equation, $\nabla_\mu F^{\mu\nu\rho\lambda} = 0$,
has the solution,
\begin{equation}
F^{\mu\nu\rho\lambda} = \frac{c}{i \sqrt{g}}
\epsilon^{\mu\nu\rho\lambda},
\label{eq-sol-F}
\end{equation}
so that $F^2 = -24 c^2$.  For real $c$ the four-form field will be
real in the Lorentzian sector (as it should) and $F^2$ will be
negative everywhere. Using this solution, it is easy to
show~\cite{DunJen90,HawTur98c} that both the dynamics and the
Euclidean action of the system can be reproduced from an action in
which all terms related to $F$ are dropped, and instead the effective
potential is substituted as follows:
\begin{equation}
V(\phi) \rightarrow U(\phi) = V(\phi) - \frac{1}{2} c^2.
\end{equation}
 
In order to explain the smallness of the observed effective
cosmological constant, $8 \pi G\, U_0$, one can make an anthropic
argument~\cite{HawTur98c} to show that the $c^2$ term will cancel the
vacuum energy remaining after the end of inflation ($\phi=0$) almost
exactly:
\begin{equation}
c^2 = 2 V_0,
\label{eq-cv}
\end{equation}
where $V_0 = V(0)$. Since $c$ is real, we must assume that $V_0 > 0$
for this cancellation to work. (Note that Brown and
Teitelboim~\cite{BroTei87,BroTei88} assume $V_0 < 0$ instead since
they choose the opposite sign for the action of the four-form field.)

While the anthropic argument works well for fixing the effective
cosmological constant, it predicts $\Omega_0 = \Omega_{\rm matter} +
\Omega_{\Lambda} = 0.01$ if the Hartle-Hawking no-boundary
proposal~\cite{HarHaw83} is used to determine the a priori probability
distribution. At the end of this paper, we will sketch a new model of
eternal inflation which avoids this problem and does not rely on the
anthropic principle to explain regions of low effective cosmological
constant.

\subsection{Open Inflation}
\label{sec-open}

In an $O(4)$-invariant Euclidean spacetime with the metric
\begin{equation}\label{metric}
ds^2 =d\sigma^2 +b^2(\sigma)(d \psi^2+ \sin^2 \psi \, d \Omega_2^2) \ ,
\end{equation}
the scalar field $\phi$ and the three-sphere radius $b$ obey the
equations of motion
\begin{equation}\label{equations}
\phi''+3{b'\over b}\phi'=U_{,\phi},~~~~~ b''= -{8\pi G\over 3} b (
\phi'^2 +U) \ ,
\end{equation}
where primes denote derivatives with respect to $\sigma$.

With the regularity conditions $b=0$, $b'=1$, and $\phi'=0$ at
$\sigma=0$ (the `North pole'), the standard distorted-four-sphere
solution is obtained. Except in the neighborhood of the South pole, it
is well approximated by an exact $S^4$,
\begin{equation}
b(\sigma) = H^{-1} \sin H \sigma,~~~~~~
\phi(\sigma) = \phi_0,
\label{eq-solution}
\end{equation}
where $H^2 = 8\pi G U(\phi_0)/3$. Near the South pole, at
$\sigma=\sigma_{\rm f}$, this approximation breaks down. The inflaton
field diverges logarithmically, while the scale factor behaves like
$(\sigma_{\rm f} - \sigma)^{1/3}$~\cite{HawTur98}.

The solution can be cut in half along the line $\psi=\pi/2$, which
removes half of each three-sphere. The resulting instanton has a
boundary of vanishing second fundamental form. Across this surface,
one can continue analytically to a Lorentzian
spacetime~\cite{GutWei83,HawTur98} with the time variable $\tau$,
given by $\psi=\pi/2+i\tau$. Thus one obtains a geodesically
incomplete part of a Lorentzian spacetime, `region II' (see
Fig.~\ref{fig-regions}):
\begin{figure}[htb]
\hspace{.2\textwidth} \vbox{\epsfxsize=.6\textwidth
\epsfbox{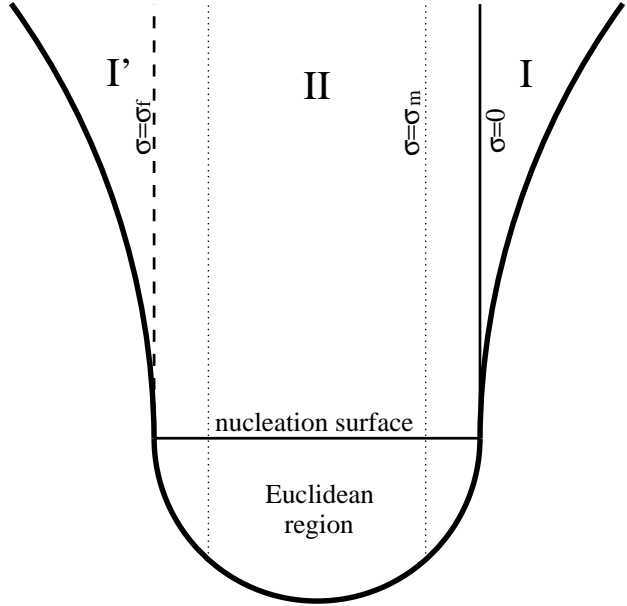}}
\caption%
{\small\sl The creation of an open inflationary universe. The light
  cone $\sigma=0$ contains the open region (I). It is surrounded by an
  exponentially expanding shell of width $H^{-1}$ (II). In the
  original HT model there is a timelike bubble singularity at
  $\sigma_{\rm f}$ which expands nearly at the speed of light; it is
  indicated by a dashed line. If this singularity is removed, for
  example by coupling the inflaton to a membrane, $\sigma_{\rm f}$ is
  another null surface surrounding a second region of open inflation
  (I$'$). To obtain a membrane, cut and identify the dotted lines.}
\label{fig-regions}
\end{figure}
\begin{equation}
ds^2 = -b^2(\sigma)\ d\tau^2 + d\sigma^2 + b^2(\sigma) \cosh^2 \tau
d\Omega_2^2.
\end{equation}
The field $\phi$ will be independent of $\tau$.  This metric describes
an exponentially expanding shell of width $H^{-1}$, which is spatially
nearly homogeneous exept in a small region near $\sigma_{\rm f}$ where
it contains a timelike singularity.

Region II is only a part of a de~Sitter-like universe. To obtain the
remainder of the manifold, one can continue across the null
hypersurface $\sigma=0$ by taking $\tau = i\pi/2 + \chi$ and $\sigma =
it$. This yields the metric
\begin{equation}
ds^2 = -dt^2 + a^2(t) \left( d\chi^2 + \sinh^2 \chi d\Omega_2^2
\right),
\end{equation}
where $ a(t) = -i\, b[\sigma(t)] \approx H^{-1} \sinh Ht$. Its
spacelike sections (defined by the hypersurfaces of constant inflaton
field) are open. The part of the spacetime covered by these
coordinates will be called region I. It looks from the inside like an
infinite open universe, which inflates while the field $\phi$ slowly
rolls down to the minimum of its potential. When the vacuum energy
ceases to dominate the evolution, the universe undergoes a transition
to a radiation or matter-dominated open Friedman-Robertson-Walker
model.

The real part of the Euclidean action of this solution comes entirely
from the Euclidean sector, and is given by
\begin{equation}
I_{\rm HT} =
      - U(\phi_0) \int_M d^4\!x\, \sqrt{g} = - \frac{\pi}{G H^2}.
\label{eq-action-HT}
\end{equation}

\subsection{Singularity problem}

The singularity occuring at the South pole gives cause for concern.
The divergence of the inflaton does not cause the volume term of the
action to diverge, and the boundary term coming from the missing point
is typically negligible. Nevertheless, singularities are not usually
allowed in gravitational instantons~\cite{EQG}. In particular, the
recent work of Vilenkin \cite{Vil98} would seem to imply that the
singularity occurring in the HT model should be disallowed
because a similar instanton would make flat space unstable.

\begin{figure}[htb]
\hspace{.34\textwidth} \vbox{\epsfxsize=.32\textwidth
\epsfbox{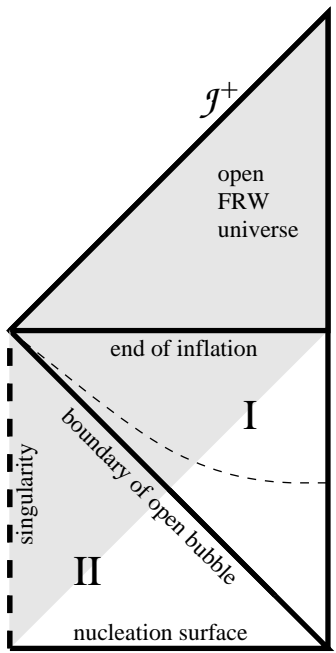}}
\caption%
{\small\sl The causal structure of the HT solution. The thin dashed
  line is a typical spacelike section in the region I of open
  inflation. While the singularity never enters the open universe,
  this Penrose diagram shows that it can be seen by observers located
  there. The shaded region is the causal future of the singularity. It
  overlaps both with the open inflationary region (I) and with the
  post-inflationary open universe. While the lines corresponding to
  the singularity and the nucleation surface in this diagram may be
  distorted, reducing the region of overlap, the singularity will
  nevertheless be visible to all post-inflationary observers at
  sufficiently late times.}
\label{fig-penrose-ht}
\end{figure}

Hawking and Turok~\cite{HawTur98c} responded by arguing that, unlike
in the flat-space case, the singularity would remain forever outside
region I, the open bubble, where observations are made. It is clear
from a causal diagram (Fig.~\ref{fig-penrose-ht}), however, that the
singularity is visible to observers in the open universe. This makes
it necessary to search for non-singular solutions leading to open
inflation.

Regular instantons have been constructed in Ref.~\cite{BouLin98} using
solutions related to those of Coleman and De~Luccia, and in
Ref.~\cite{Gon98b} using axionic wormholes. Garriga argued in
Ref.~\cite{Gar98b} that the singularity may disappear for compactified
five-dimensional Kaluza-Klein instantons, at the expense of some
fine-tuning. In Ref.~\cite{Gar98a} by the same author, the singularity
is cut out using domain walls of negative energy density.

It seems to us, however, that allowing negative energy is just as
problematic as allowing singularities. Therefore, we will consider
only domain walls, or membranes, with strictly positive energy
density. The divergence of the inflaton field is averted by coupling
it to the domain wall. The advantage of this approach is that no
particular features in the inflaton potential need to be assumed. Like
the original HT model, it works for generic models of
chaotic inflation, but without the problem of a singularity.

Introducing a membrane into the model is no random fix. The membrane
of supergravity is the natural source for the four-form field which is
already in the theory. It ought to be included in any case. Such
membranes are characterized by a mass and a charge. In
Sec.~\ref{sec-domainwall} we construct a non-singular instanton
containing an uncharged membranes (a domain wall). It is obtained as a
regular portion of the HT instanton, glued to its own mirror image.
This instanton describes the nucleation of a universe containing two
regions of open inflation, separated by a domain wall. Alternatively,
it may be used to describe the spontaneous nucleation of a membrane in
a pre-existing de~Sitter background.

In Sec.~\ref{sec-membrane} we consider charged membranes, across which
the four-form field, and thus the effective cosmological constant, has
a discontinuity. We show how to obtain an instanton containing a
membrane by gluing portions of two different HT instantons together.
Finally, we outline a new model of eternal inflation, driven by the
four-form field, in which the membrane instanton plays a crucial role.

\sect{Domain wall wrapping an open universe}
\label{sec-domainwall}

\subsection{Domain walls}

In cosmology, the term ``domain wall'' normally refers to
two-dimensional topological defects which can form whenever there is a
breaking of a discrete symmetry.  Commonly, one thinks of the symmetry
breaking in terms of some Higgs field, $\Phi$.  If ${\cal M}_{0}$
denotes the vacuum manifold of $\Phi$ (i.e., the submanifold of the
Higgs field configuration space on which the Higgs acquires a vacuum
expectation value because it will minimize the potential energy
$V(\Phi)$), then a necessary condition for a domain wall to exist is
that ${\pi}_{0}({\cal M}_{0}) \not= 0$.  In other words, vacuum domain
walls arise whenever the vacuum manifold is not connected.  The
simplest example of a potential which gives rise to vacuum
domain walls is the classic `double well', which is
discussed in detail (along with many related things) in
Ref.~\cite{VilShe94}.

In general, domain walls are $(D-2)$-dimensional defects (or extended
objects) in $D$-dimensional spacetimes.  As such, they are a common
feature in the menagerie of objects which arise in the low-energy
limit of string theory, as has been discussed in detail in
Refs.~\cite{CowLu97} and \cite{CveSol96}.  In other words, in string
theory a domain wall is just another `brane in the crowd'.  In this
case they will usually carry a charge. For $D=4$, such membranes act
as a source for the four-form field, and change its value. We will
investigate the effects of charge in Sec.~\ref{sec-membrane}. It is
instructive, however, to consider first an uncharged domain wall,
which is characterized, in the thin wall approximation, only by its
energy density, $\mu$.

Its Euclidean action is given by
\begin{equation}
I_{\rm dw} = \mu \int d^3\!x\, \sqrt{h},
\end{equation}
where the integral is over the domain wall world volume, and $h =
\det{h_{ij}}$ refers to the induced three-metric.
The energy density of the domain wall is related to a discontinuity in
the extrinsic curvature, $K_{ij}$, via the Israel matching conditions:
\begin{enumerate}
\item 
A domain wall hypersurface is totally umbilic, i.e., the second
fundamental form $K_{ij}$ is proportional to the
induced metric $h_{ij}$ on each domain wall world sheet.
\item The discontinuity in the second fundamental form on each domain
  wall hypersurface is $K_{ij}^+ - K_{ij}^- = 4 \pi G \mu h_{ij}$,
\end{enumerate}
where $K_{ij}^{-}$ ($K_{ij}^{+}$) refers to the limiting value of
the extrinsic curvature on the side of the wall to which the normal
vector is (not) pointing. These conditions, together with the
Gibbons-Hawking boundary terms evaluated on the wall, yield the
on-shell action
\begin{equation}
I_{\rm DW} = - \frac{\mu}{2} \int d^3\!x\, \sqrt{h};
\label{eq-action-DW}
\end{equation}
the volume integrals in the gravitational action,
Eq.~(\ref{eq-action-ht}), are now understood to exclude the domain
wall.

\subsection{Cosmological domain wall instantons}

In the instanton used by Hawking and Turok for the nucleation of an
open inflationary universe, Eq.~(\ref{eq-solution}), the hypersurfaces
of constant $\sigma$ are trivially umbilic; we shall build our domain
walls there. When the analytic continuation to a Lorentzian spacetime
is performed, these surfaces remain forever just outside the region of
open inflation, draping it but not intruding.

\begin{figure}[htb]
\epsfysize=\textwidth
\rotate[r]{\epsfbox{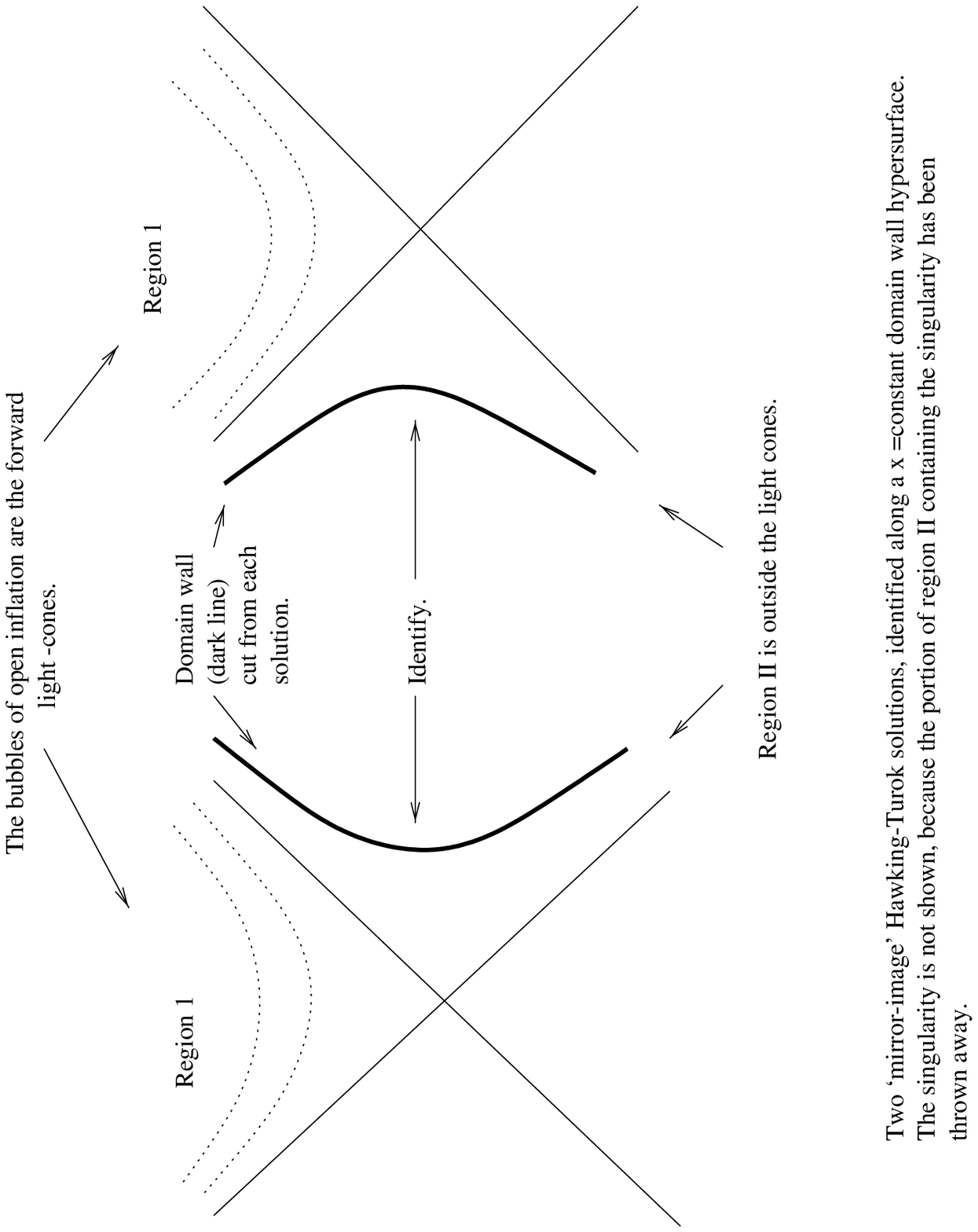}}
\caption[Two open inflating universes.]
{\small\sl Cut, copy and paste: How to construct a solution containing
  two regions of open inflation separated by a domain wall.}
\label{fig-cp1}
\end{figure}

The idea is to cut the instanton along one of these surfaces and
discard the singular part. The remaining part of the distorted
four-sphere is then glued onto its mirror image. In the Lorentzian
sector, this means we get rid of the part containing the singularity
at ${\sigma}_{\rm f}$, but keep the part containing region I, the open
universe; in fact we will obtain two such regions (see
Fig.~\ref{fig-cp1}).
This `cut, copy and paste' procedure gives rise to a discontinuity in
the second fundamental form, and thus to a domain wall. It also leads
to a small discontinuity in the derivative of the inflaton field,
$\phi$; we will discuss in Sec.~\ref{sec-coupling} how this can be
attributed to an interaction of the field with the membrane.  For now,
we shall concentrate on the geometric aspects, using the approximation
of constant $\phi$ throughout the instanton.

\begin{figure}[htb]
  \hspace{.3\textwidth} \vbox{\epsfysize=.4\textwidth
\rotate[r]{\epsfbox{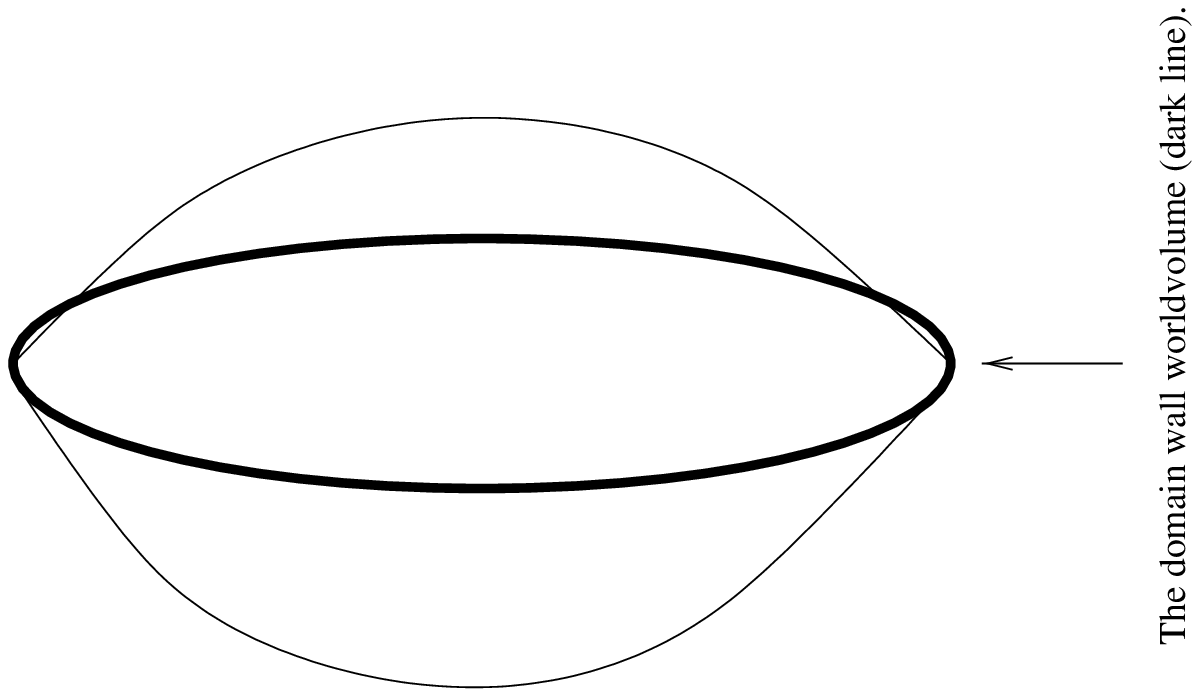}}}
\caption[The `hamburger instanton' is obtained by gluing together two
portions of the HT instanton.]
{\small\sl The hamburger instanton is obtained by gluing a portion
  of the $S^4$ instanton to its mirror image.}
\label{fig-ham}
\end{figure}

We thus take the portion of the four-sphere, Eq.~(\ref{eq-solution}),
corresponding to $0 \leq \sigma < \sigma_{\rm m}$ and match it to its
mirror image.  The extrinsic curvature at $\sigma = \sigma_{\rm m}$
will be given by
\begin{equation}
K_{ij}^{\pm} =
 {\pm}\frac{b^{\prime}(\sigma_{\rm m})}{b(\sigma_{\rm m})}{h_{ij}},
\end{equation}
where the prime denotes differentiation relative to $\sigma$.
According to the Israel conditions, this gives rise to a domain wall
of energy density $\mu$:
\begin{equation}
{K_{ij}}^{+} - {K_{ij}}^{-} = 4{\pi}G{\mu}h_{ij},
\end{equation}
whence
\begin{equation}
{\mu} = \frac{b^{\prime}(\sigma_{\rm m})}{2{\pi}Gb(\sigma_{\rm m})}.
\label{eq-sigma-b}
\end{equation}
We are assuming, of course, that $\mu$ is a fixed
parameter in the theory. Then the above equation tells us where to
cut:
\begin{equation}
\sigma_{\rm m} = H^{-1} {\rm arccot}\, 2 \pi G \mu H^{-1} .
\end{equation}
In particular, for positive energy domain walls, we find that
$b^{\prime} > 0$. In other words, one must cut before the surface of
maximum expansion, $\sigma<\sigma_{\rm max}$, and discard the larger
part of the instanton. Gluing the remaining $\sigma<\sigma_{\rm m}$
sector of the $S^4$ instanton to its mirror image, we thus obtain an
instanton shaped like a lens, or perhaps a hamburger, with the domain
wall playing the role of the beef (see Fig.~\ref{fig-ham}).

The corresponding Lorentzian solution is obtained by the same
Coleman-De~Luccia analytic continuations that were employed for the
HT instanton, Eq.~(\ref{eq-solution}). The fact that a part
of the four-sphere is missing means that the Lorentzian shell
surrounding the open universe (`region II') will be correspondingly
thinner than the usual $H^{-1}$. The domain wall resides in the center
of this shell, at constant $\sigma$.

The Euclidean action can easily be calculated from the on-shell
actions for the $S^4$ instanton, Eq.~(\ref{eq-action-HT}), and the
domain wall, Eq.~(\ref{eq-action-DW}):
\begin{equation}
I_{\rm HT/D} = - U(\phi_0) \int_M d^4\!x\, \sqrt{g}
               - \frac{\mu}{2} \int d^3\!x\, \sqrt{h}.
\end{equation}
Adjusting for the smaller four-volume that arises from restricting to
$\sigma<\sigma_{\rm m}$ on each half of the hamburger, and noting that
the Euclidean world volume of the domain wall is a three-sphere of
radius $b(\sigma_{\rm m})$, one obtains
\begin{equation}
I_{\rm HT/D} = - \frac{\pi}{G H^2} (1- \cos H\sigma_{\rm m}).
\end{equation}

As we discussed in the introduction, the hamburger instanton can be
applied in two ways. It may be interpreted to describe the
semiclassical creation from nothing of two open universes surrounded
and separated by a membrane. Alternatively, one can use it to describe
the spontaneous nucleation of a domain wall in a pre-existing
de~Sitter-like universe.  To obtain the semiclassical creation rate,
one must first subtract the background action,
Eq.~(\ref{eq-action-HT}). This yields
\begin{equation}
\Gamma = e^{-(I_{\rm HT/D} - I_{\rm HT})}
   = \exp \left(- \frac{\pi}{G H^2} \cos H\sigma_{\rm m} \right).
\end{equation}

\subsection{Using domain walls to avoid the singularity}
\label{sec-coupling}

The domain wall we have introduced effectively cuts the
distorted-four-sphere instanton, discards the larger (singular) part,
and pastes the smaller part to its mirror image to form the hamburger
instanton. The two parts are not completely symmetric, however, since
the inflaton field, $\phi$, grows slowly with Euclidean time, and
continues to grow across the domain wall. It would therefore diverge
near the South pole in the same way that it does in the HT
case, leading to a mild singularity.

In the presence of a domain wall, however, it is easy to prevent this
behavior. All it takes is a small coupling of the domain wall to the
field:
\begin{equation}
\mu = \mu(\phi).
\end{equation}
Garriga~\cite{Gar98a} used the coupling $\mu = \mu_0 - \alpha
e^{\kappa \phi}$ in a different context, but for our purposes there is
no need to assume any specific form. With the $\phi$-dependence of
$\mu$, the inflaton field moves in the effective potential
\begin{equation}
U_\mu(\phi) = U(\phi) + \delta(\sigma - \sigma_{\rm m}) \mu(\phi).
\end{equation}
This leads to a discontinuity of $\phi'$ across the domain wall:
\begin{equation}
\phi'_- - \phi'_+ = \mu_{,\phi}.
\end{equation}
The instanton will be non-singular if it is made completely symmetric,
i.e. if $\phi'_- = - \phi'_+$. Thus we obtain the regularity condition
\begin{equation}
\mu_{,\phi}[\phi(\sigma_{\rm m})] = -2 \phi'(\sigma_{\rm m}).
\end{equation}
This does, of course, require fine-tuning of the coupling. We can turn
the argument around, however, by noting that $\mu_{,\phi}$ generically
depends on the value of $\phi$ at the domain wall, which is
approximately $\phi_0$. Requiring regularity for a given coupling thus
fixes the initial value of the inflaton field. Therefore, the coupling
can be used to dial $\phi_0$, and thus the value of $\Omega_0$.

\sect{Charged membrane wrapping an open universe}
\label{sec-membrane}

\subsection{Fundamental Membranes}

A feature of the $p$-branes of supergravity theories is that they are
naturally associated with $(p+2)$ forms, which may be regarded as the
field strengths for $(p+1)$ potentials. The potential couples
electrically to the worldvolume of the $p$-brane.  Thus, by
Eq.~(\ref{eq-fa}), the four-form field $F_{\mu\nu\rho\lambda}$ is
related to the three-form potential $A_{\nu\rho\lambda}$, which
couples to the worldvolume of a membrane, i.e.\ a two-dimensional
extended object.  In four spacetime dimensions, one would regard a
two-dimensional extended object as a `domain wall'.  Put the other way
around, one can couple any domain wall to the four-form in this way.
Such a coupling was considered in a cosmological context in
\cite{BroTei87,BroTei88}, where it was shown that the effective
cosmological constant can be decreased through the nucleation of
membranes which couple to the four-form.

In string theory there are in general three distinct classes of
$p$-branes, which are distinguished according to how the tension of
the brane varies with the string coupling constant.  Let $g_{\rm s}$
denote the string coupling constant, which, crudely speaking, measures
the size of the eleventh dimension in M-theory.  Then the three
classes, consisting of fundamental branes, Dirichlet branes
(D-branes), and solitonic branes, may be defined as follows
\cite{Sch97}.  For fundamental $p$-branes, the tension $T_p$ does not
depend at all on the string coupling.  For D$(p)$-branes, the tension
$T_p$ varies inversely as the string coupling ($T_{p} ~{\sim}~
1/g_{\rm s}$).  Finally, for solitonic branes the tension varies
inversely as the square of the coupling ($T_{p} ~{\sim}~ 1/g_{\rm
  s}^2$).

The vacuum expectation value of the dilaton field, $g_{\rm s}$, may
not be constant in general.  If the effective energy density of a
domain wall depends on a field which is spacetime dependent, one may
not use the Israel conditions, since these conditions are derived
subject to the assumption that the energy density of of the wall is
constant.  Thus, in order to describe D8-branes (in ten dimensions)
which are domain walls
in backgrounds with a varying dilaton, for example, one needs to
generalize the Israel conditions in some way.  The correct
modification of the Israel conditions will be presented in a future
paper~\cite{ChaPer98}.

Here, however, we are going to assume that the domain walls couple as
{\it fundamental} branes to the four-form.  This assumption, which was
implicitly made in \cite{BroTei87,BroTei88}, is important, because it
means that we do not have to worry about any dilatonic dependence for
the energy density of the domain walls, and so we are justified in
using the Israel conditions to construct domain wall hypersurfaces.

With this in mind, we follow the constructions in
Ref.~\cite{BroTei87,BroTei88} and couple our domain walls to the
four-form of \cite{HawTur98c}.  In Ref.~\cite{HawTur98c}, the
four-form was introduced in an attempt to `solve' the cosmological
constant problem.  Dirac quantization for the charges supported by
these branes~\cite{Sch97} implies that the effective cosmological
constant generated in this way will jump by an integral multiple of a
discrete unit of charge, as one moves from one side of the membrane to
the other side.  This does not necessarily imply a fine-tuning problem
for the four-form of HT~\cite{HawTur98c}, simply because their
four-form is not induced by a membrane.  This is analagous to the
situation in ordinary electromagnetism, where it only makes sense to
talk about charge quantization once electric sources have been
introduced.  In the absence of elementary particles, electric flux can
take any value.

\subsection{Action and Solutions}

The Euclidean action of a charged membrane is given by
\begin{equation}
I_{\rm dw} = \mu \int d^3\!x\, \sqrt{h}
   + \frac{e}{6} \int d^3\!x\, A_{\nu\rho\lambda}
                        \epsilon^{\nu\rho\lambda}
\end{equation}
The solution for the four-form field will still be given by
Eq.~(\ref{eq-sol-F}), except that the new $\delta$-function term in
its equation of motion leads to a discontinuity of the field strength
across the domain wall:
\begin{equation}
F^{\mu\nu\rho\lambda} = \frac{c_\pm}{i \sqrt{g}}
\epsilon^{\mu\nu\rho\lambda},
\label{eq-sol-F2}
\end{equation}
with
\begin{equation}
c_- = c_+ - e.
\label{eq-jump}
\end{equation}
As before, the super- or subscript `$+$' refers to the region before
the membrane ($\sigma < \sigma_{\rm m}$), and `$-$' to the region
beyond. The instanton will consist of sections of two distorted
four-spheres of different radii, matched across the membrane:
\begin{eqnarray}
b(\sigma) & = & H_+^{-1}
 \sin H_+ \sigma~~~~(\sigma < \sigma_{\rm m}),
\\
b(\sigma) & = & H_-^{-1}
 \sin H_- (\sigma_{\rm f} - \sigma)~~~~(\sigma > \sigma_{\rm m}),
\end{eqnarray}
where
\begin{equation}
H_\pm^2 = 8\pi G U_\pm(\phi_0)/3,
\end{equation}
and
\begin{equation}
U_\pm(\phi) = V(\phi) - \frac{1}{2} c_\pm^2.
\label{eq-Upm}
\end{equation}
We use the approximation $\phi(\sigma) = \phi_0$. It is
straightforward to generalize the method used in
Sec.~\ref{sec-coupling} to remove the singularity in the charged case;
we will therefore not worry about it here. Also, the analytic
continuation to a Lorentzian spacetime is exactly as for the HT
instanton without domain walls, or with uncharged domain walls (see
Sec.~\ref{sec-open}).

The matching analysis is more interesting now. As the solution is
regular, there will be two open universes, emerging from light cones
at $\sigma=0$ and $\sigma = \sigma_{\rm f}$. They will differ in the
value of the cosmological constant after inflation ends, $U_0 = V_0 -
\frac{1}{2} c_\pm^2$. In order for the universe to live long enough,
and for galaxies to form, it is necessary that $U_0 \approx 0$ in one
of the two open regions. We may choose this region to be on the `$+$'
side. As we restrict to compact instantons, we shall not consider
Anti-de~Sitter spacetimes. Therefore we will take $e>0$, so that the
`$-$' bubble will retain a higher residual cosmological constant. Then
the effective cosmological constant in the `$+$' portion of the
instanton will be lower, and the radius of curvature larger, than in
the `$-$' portion. From the positivity of the membrane energy, it then
follows that the `$+$' portion will always be less than half of an
$S^4$. The `$-$' portion, however, can be either less ($\eta = 1$) or
more ($\eta = -1$) than half.

With $c_+$ fixed by anthropic arguments, the values of $H_\pm$ depend
on $\phi_0$ and $e$ according to Eqs.~(\ref{eq-jump}) and
(\ref{eq-Upm}). The energy density, $\mu$, and the requirement that
the scale factor be continuous across the membrane fix $\sigma_{\rm
  m}$ and $\sigma_{\rm f}$. It was convenient in the previous section
to eliminate $\mu$ in favor of $\sigma_{\rm m}$; similarly, it will
simplify expressions now if we eliminate in favor of $b_{\rm m} =
b(\sigma_{\rm m})$, which is related to $\mu$ by
\begin{equation}
\mu = \frac{ \sqrt{1-H_+^2 b_{\rm m}^2} + \eta
  \sqrt{1-H_-^2 b_{\rm m}^2}}{4 \pi G b_{\rm m}}
\end{equation}
Thus we shall specify instantons by $(H_+, H_-, b_{\rm m}, \eta)$
instead of the equivalent $(\phi_0, e, \mu)$.

On-shell the charge term vanishes and the instanton action is given by
\begin{equation}
I_{\rm HT/M} = -\frac{\pi}{2 G H_-^2} \left( 1 - \eta
 \sqrt{1-H_-^2 b_{\rm m}^2} \right)
-\frac{\pi}{2 G H_+^2} \left( 1 -
 \sqrt{1-H_+^2 b_{\rm m}^2} \right).
\end{equation}
This is the relevant action if we consider the creation from nothing
of a universe containing two bubbles of open inflation with different
cosmological constants, separated by a charged membrane.

If we are interested in the spontaneous nucleation of a charged
membrane on a background de~Sitter-like universe, we must subtract the
action of the background instanton to get the pair creation rate.
There are two possible backgrounds: the four-spheres of radius
$H_\pm^{-1}$. Using the `$+$' background corresponds to the nucleation
of a membrane inside of which the cosmological constant is larger. By
Eqs.~(\ref{eq-pcr}) and (\ref{eq-action-HT}), the creation rate is
$\exp ( - \Delta I^+_{\rm HT/M} )$, with
\begin{equation}
\Delta I^+_{\rm HT/M} = \frac{\pi}{2 G H_+^2} \left( 1 +
 \sqrt{1-H_+^2 b_{\rm m}^2} \right)
-\frac{\pi}{2 G H_-^2} \left( 1 - \eta
 \sqrt{1-H_-^2 b_{\rm m}^2} \right).
\end{equation}
Starting from the `$-$' background, on the other hand, means that the
membrane nucleation decreases the cosmological constant. This occurs
at a rate of $\exp ( - \Delta I^-_{\rm HT/M} )$, with
\begin{equation}
\Delta I^-_{\rm HT/M} = \frac{\pi}{2 G H_-^2} \left( 1 + \eta
 \sqrt{1-H_-^2 b_{\rm m}^2} \right)
-\frac{\pi}{2 G H_+^2} \left( 1 -
 \sqrt{1-H_+^2 b_{\rm m}^2} \right).
\end{equation}
Starting from a given background, a little algebra shows that either
type of nucleation is suppressed, and that the increase of the
effective cosmological constant is more suppressed than the decrease:
\begin{equation}
\Delta I^+_{\rm HT/M} > \Delta I^-_{\rm HT/M} > 0.
\end{equation}

Thus, the cosmological constant is, on average, driven down by
membrane creation.  For an appropriate range of parameters, it can
relax to a value within experimental bounds, and will not decrease
below zero~\cite{BroTei87,BroTei88}. The problems with this kind of
mechanism are two-fold. First, why should the unit of membrane charge
be so small as to be able to tune the cosmological constant to an
accuracy of $10^{-122}$? It is simply unclear where such a scale
should come from. However, this problem is still less serious than the
original cosmological constant problem; once the scale is explained,
dynamic fine-tuning takes place automatically.

As was pointed out in Refs.~\cite{BroTei87,BroTei88}, however, the
main problem is the slowness of the process. By the time the final
jump of the effective cosmological constant occurs, the inflaton field
will be in the minimum of its potential, and the open universe
contained in the membrane will be empty. There is one way to
circumvent this problem, which we will turn to next. (We should also
point out that Linde~\cite{Lin98} is proposing a non-dynamical fix for
the cosmological constant in the wavefunction framework. In this
scenario, the anthropic principle selects an acceptable value for the
cosmological constant just like in Ref.~\cite{HawTur98c}, but the
problem of low $\Omega_0$ is avoided.)

\subsection{Eternal inflation driven by the four-form}
\label{sec-eternal}

Let us assume that the universe is in a period of inflation, with
$\phi$ far from its minimum and slowly rolling down; also assume that
$U_0 = V_0-\frac{1}{2} c^2$ has not been fine-tuned and is many orders
of magnitude larger than current constraints allow. The cosmological
constant would need to be corrected by membrane creation at least $60$
e-foldings before inflation ends in order to obtain a realistic value
of $\Omega_0$.  Given the exponential suppression of membrane
nucleation events, they are extremely unlikely to take place even
during the entire classical roll-down of the inflaton field. The need
to complete the dynamic tuning, rather than just achieve some random
decrease of the cosmological constant, exacerbates this problem.

Linde has pointed out~\cite{Lin86a} that for generic models of chaotic
inflation, there is a critical value of the effective cosmological
constant beyond which random quantum fluctuations of the inflaton
field dominate over the classical decrease. As the number of horizon
volumes grows exponentially, the effective cosmological constant will
increase roughly in half of the new domains, so that inflation
continues forever in some regions.

If the cosmological constant arises purely from the inflaton potential
(i.e. if $V_0=c=0$), the regime of eternal inflation lies beyond a
critical value of $\phi$, for which de~Sitter space is hot enough to
support the strong quantum fluctuations. The exit from the eternal
phase occurs locally if $\phi$ jumps below the critical value. This
leads a region of spacetime into the regime of classical slow-roll,
and allows inflation to end there. The entire observable universe
would be contained in such a region.

In models with a four-form field, there is a different way to support
eternal inflation: If $U_0$ is sufficiently large, the temperature of
the spacetime, $T \sim H/2\pi$, will support a stochastic quantum
evolution of the inflaton field for all values of $\phi$, even
$\phi=0$. In this eternally inflating universe, all values of $\phi$
will be realized in different regions, from $\phi=0$ to $\phi =
\phi_{\rm Pl}$. Then the number of horizon volumes is unbounded, and
the suppression of membrane creation is no longer an obstacle. In
fact, the exit from this type of eternal inflation occurs not though a
jump of $\phi$, but instead by membrane nucleation pushing $U_0$ below
the critical value.

Because of the stochastic distribution of the values of $\phi$, more
than $60$ e-foldings of inflation will occur inside most membranes.
For sufficiently high values of $\phi$, the transition may simply be
one from $F^2$-driven eternal inflation to `ordinary' eternal
inflation. For lower values of $\phi$, the classical decrease
immediately begins. This makes no difference; the only thing that
matters is that the exit sets $U_0 \approx 0$.

Most membranes that are spontaneously created will fail to do so,
because they will have the `wrong' charge. They may trigger a local
exit from eternal inflation, but without leading to an acceptable
value of the present cosmological constant.  We do not worry about
these membranes; they merely add to the vast vacuum-dominated regions
abundant in eternal inflation.  However, a small proportion of
membranes will have just the right charge (some appropriate multiple
of the unit charge), such that the cosmological constant is
neutralized, or nearly enough neutralized, by a single nucleation
event. In an eternally inflating universe, any small but
semiclassically non-vanishing probability is large
enough~\cite{Bou98}. There will be regions in which inflation ends and
galaxies can form, typically with $\Omega_{\rm matter}+\Omega_\Lambda
= 1$, although smaller values of $\Omega_0$ will also occur. That we
live in such a region is no more surprising than the fact that we do
not live in intergalactic space.

Such models may still suffer from some of standard problems that come
with combining a fundamental eleven-dimensional theory with
inflationary theory: the inflaton potential must be inserted ad hoc,
and there is no convincing argument why the unit membrane charge, and
the background value of $F^2$, should not both be of Planckian order.
But it has several strong advantages over other proposals. First, like
in all models of eternal inflation, any initial conditions are
completely obliterated, allowing us to avoid the minefield of
wavefunction proposals. Second, as in Refs.~\cite{BroTei87,BroTei88},
the effective cosmological constant is neutralized dynamically, with
no need to resort to anthropic arguments.  Third, it predicts a total
value of $\Omega_0=1$, which is consistent with recent supernova
observations.

\sect{Summary}
\label{sec-discussion}

We have shown that it is possible to couple domain walls to the
four-form field recently resuscitated by Hawking and Turok to `solve'
the cosmological constant problem, and that one may regard such
membranes as sources for the effective cosmological constant. For
generic inflaton potentials, we constructed non-singular instantons
that give rise to a Lorentzian universe containing two regions of open
inflation, separated by a charged or uncharged membrane.

Such instantons may be interpreted to describe the creation of
singularity-free, open inflationary universes from nothing.  They also
mediate the semiclassical production of membranes on a pre-existing
inflating background; on average, this process drives down the
effective cosmological constant. We proposed a model of eternal
inflation in which regions with acceptable values of $\Omega_0$ and
$\Omega_\Lambda$ are produced dynamically.

The membranes we considered wrap the open universes, hovering just
outside the light-cones that bound the regions of open inflation.
Parts of the inflationary and postinflationary open universe are in
their causal future. Fluctuations generated by perturbing these domain
walls travel into the bubble of open inflation.  It would be
interesting to understand whether they can have observable effects.

\subsection*{Acknowledgements}

We would like to thank our colleagues in Cambridge and Stanford,
especially Gary Gibbons, Stephen Hawking, Nemanja Kaloper, and Andrei
Linde, for discussions. We are particularly grateful to Andrei Linde
for comments on a draft of this paper. R.B.\ was supported by
NATO/DAAD.  A.C.\ was supported by Pembroke College, Cambridge.

\bibliographystyle{board}
\bibliography{all}

\end{document}